\documentstyle[epsfig,12pt]{article} 
  
\newcommand{\fig}[1]{Fig.\ref{#1}}
\newcommand{\tab}[1]{table \ref{#1}}

\newcommand{\eqn}[1]{Eq.(\ref{#1})}
 
\newcommand{\bsm}{\begin{small}} 
\newcommand{\esm}{\end{small}} 
\newcommand{\bc}{\begin{center}} 
\newcommand{\ec}{\end{center}} 
\newcommand{\ben}{\begin{enumerate}} 
\newcommand{\een}{\end{enumerate}} 
\newcommand{\bq}{\begin{equation}} 
\newcommand{\eq}{\end{equation}} 
\newcommand{\bqa}{\begin{eqnarray}} 
\newcommand{\eqa}{\end{eqnarray}} 
\newcommand{\nn}{\nonumber} 
 
\def\demo{$\Delta\eta\mu \acute{o} \kappa \varrho \iota \tau o \varsigma$} 
\def\onehalf{\frac{1}{2} }

\begin{document} 
\begin{titlepage} 

${}$ \hfill CERN-TH/98-207 \\
${}$ \hfill DEMO-HEP.98/01

 
\vspace*{2cm} 

\bc 
\begin{LARGE} 
{\bf  
On the computation of multigluon amplitudes
}\\ \end{LARGE} 
\vspace*{2cm} 
{\bf   
Petros Draggiotis $^a$\footnote[1]{e-mail: pdrag@alice.nrcps.ariadne-t.gr} 
Ronald H. P. Kleiss $^b$\footnote[2]{e-mail: kleiss@sci.kun.nl}
Costas G. Papadopoulos$^{ac}$\footnote[3]{e-mail: 
Costas.Papadopoulos@cern.ch} }\\[12pt]

$^a$Institute of Nuclear Physics, NCSR \demo, 15310, Athens, Greece \\[6pt] 
$^b$University of Nijmegen, The Netherlands \\[6pt]  
$^c$CERN, Theory Division, CH-1211, Geneva, Switzerland   
\vspace*{1cm} 
\begin{abstract} 
 
A computational algorithm based on recursive equations is developed 
in order to estimate multigluon production processes at high energy
hadron colliders. The partonic reactions $gg\to (n-2)\, g$ with $n\le 9$ 
are studied 
and comparisons with known approximations are presented.
\end{abstract} 
 
\ec 
 
\vspace*{\fill}
\indent 
June 1998 
\vspace*{1cm}
\end{titlepage} 
 
\newpage 
\pagestyle{plain} 
\setcounter{page}{1} 
 
\section{ Introduction }  

High-energy hadron colliders will be valuable tools for developing particle physics
in the next decade, especially with the planned run of the 
Large Hadron Collider (LHC),
which will operate
at $\sqrt{s}=14$~TeV, the highest energy achieved so far. 
Multijet final states are  of
great importance at high energies, and in fact events with very high jet 
multiplicities,
up to six jets, have already been observed and measured at the
 TEVATRON collider~\cite{TEVATRON}.

On the other hand, the estimate of the multijet production cross sections 
as well as
their characteristic distributions is a formidable task from the theoretical 
point of view~\cite{gluons}.
Perturbation theory based on Feynman graphs, the `brute-force' method, 
runs into computational problems, since the
number of graphs contributing to the $n$-gluon amplitude grows like $n!$, 
jeopardising the
computational efficiency. 
In order to simplify the computation, colour 
decomposition~\cite{Ber-Gie1,Man-Par,Man-Par-Xu}
has been used extensively
in the past. According to this scheme, the $n$-gluon amplitude 
can be written as:

\bq
{\cal M}= 2 i g^{n-2} \sum_{P(2,\ldots,n)} Tr(t^{a_1}\ldots t^{a_n}) 
{\cal C}(1,\ldots,n)
\eq
where $t^a$ are the generators of the $SU(N)$ group 
in the fundamental representation,
$g$ is the gauge coupling constant and ${\cal C}(1,\ldots,n)$ 
are gauge invariant
subamplitudes that do not contain any explicit information on the colour
structure. The computational complexity is now due to the large number
of gauge invariant subamplitudes, which is roughly proprtional to $n!$.
Moreover the colour summation leads to a colour matrix with dimensionality
$(n-2)!\times (n-2)!$~\cite{Ber-KujI}, another source of computational 
inefficiencies.

On the other hand recursive 
methods~\cite{Ber-Gie2,Ber-KujI,Ber-KujII,Arg-Kle-Pap,Car-Mor} 
have been used extensively in order to overcome the computational
obstacles of the `brute-force' method. Following this line of thinking 
we present in this 
paper results on $gg\to (n-2)\, g$ with $n$ up to $n=9$. 
The main ingredients we have used, in order
to make the computation as simple and efficient as possible, are
\begin{enumerate}
\item The amplitude is computed using 
recursive equations which result in a computational cost 
growing asymptotically as $~3^n$, to be compared with the $n!$ growth in the
`brute-force' method~\cite{Bru}.
\item Colour and helicity structure are appropriately transformed so that
a replacement of the usual summation with Monte Carlo integration is made
possible.
\end{enumerate}

\section{ The Recursive Equations for QCD }  

\subsection{ The Generating Function }

Let us assume that $\phi(x)$ represents a field with arbitrary quantum numbers.
The generating function of all connected Green functions is given by
the standard formula~\cite{Peskin}
\bq
Z[J]= -\Gamma[\phi]+\int d^4x \phi(x) J(x)\,,
\eq
subject to the condition that $\phi$ is a solution of the field equations
\bq 
\frac{\delta \Gamma}{\delta \phi} = J\,.
\eq
$\Gamma[\phi]$ is the effective action which at tree order is given by
\bq
\Gamma[\phi]=\int d^4x {\cal L}\,,
\eq
where ${\cal L}$ is the Lagrangian of the theory.
For a process with $n$ external particles the source, $J(x)$, is given as
an expansion over the free asymptotic states,
\bq
J(x)=\sum_{l=1}^{n}a_l \,e^{-i p_l\cdot x}\,,
\eq
or in momentum space by
\bq
J(q)= (2 \pi)^4 \, \sum_{l=1}^{n}a_l \,\delta(q-p_l)\,,
\eq
The standard LSZ reduction formula relates the amplitude with the $n$-th
derivative of the generating function
\bq
{\cal A}(p_1,p_2,\ldots,p_n)=\lim\prod_{l=1}^{n}[Z^{(2)}(p_l)]^{-1}
\frac{\partial^n Z}{\partial a_1 \ldots \partial a_n}
\eq
where $[Z^{(2)}(p)]^{-1}$ is the inverse propagator and the on-shell limit
is understood.
Now it is not difficult to prove that the field equations aquire the following
solution~\cite{Car-Mor,Drag}

\bq 
\phi(q)=(2 \pi)^4 \, \sum_{m=1}^{2^n-2} b_m \,\delta(q-P_m)
\label{fie-eq}
\eq
where $P$ are all possible sums of the $n$ external momenta, $p_i$, 
$i=1,\ldots,n$. For example, $p_1+p_2$, $p_1+p_4+p_n$, etc. Their number is
simply determined by the binomial sum
\[
\sum_{l=1}^{n-1} \left( \begin{array}{c} 
                          n \\
                          l
                       \end{array} \right) = 2^n-2\,.
\]

The coefficients
$b_m$ satisfy non-linear equations which can be solved iteratively leading 
to recursive equations.
Since we are interested in multigluon amplitudes, we study the derivation
of these equations in the case of QCD. The Lagrangian for QCD is given by: 
\bq 
{\cal L }={\cal L}_{0} +{\cal L}_{int} 
\eq 
\bq 
{\cal L}_{0} =- \frac{1}{4} ( \partial_{ \mu } A_{ \nu }^{a}- 
\partial_{ \nu } A_{ \mu }^{a}) (  \partial^{ \mu } A^{ \nu  a}- 
\partial^{ \nu } A^{ \mu a})- \frac{1}{2 \xi } ( \partial^{ \mu } A_{ \mu }^{a} )^2 
\eq 
\bq 
{\cal L}_{int} =- \frac{g}{2} f^{ abc } ( \partial_{ \mu } A_{ \nu }^{a}- 
\partial_{ \nu } A_{ \mu }^{a}) A^{ b \mu } A^{c \nu } 
-\frac{g^2}{4} f^{abe} f^{cde} A^{a}_{\mu} A^{b}_{\nu} A^{ c \mu } A^{ d \nu } 
\eq 
where  $\xi$ is the gauge fixing parameter. We shall take the value $\xi=1$ 
for the rest of this paper. 
In momentum space the effective action, $\Gamma=\int d^4x {\cal L}$, is written as: 
\bqa 
\Gamma& =& -\onehalf  \int \frac{d^4q}{(2\pi)^4} A^{a}_{\mu} 
(-q) \delta^{ab} g^{\mu \nu } q^2 A^{b}_{\nu} (q)    \\ \nn 
& +& \frac{i g}{3!} 
\int \left( \prod_{i=1}^{3}\frac{d^4q_i}{(2\pi)^4}\right)
V^{abc}_{\nu \kappa \lambda } (q_1, q_2 , q_3) A^{a \nu} 
(q_1) A^{b \kappa} (q_2)  
A^{c \lambda} (q_3)(2 \pi)^4 \: \delta^4 (\sum_{i=1}^3 q_i )   \\ \nn 
&-& \frac{g^2}{4!} 
\int \left( \prod_{i=1}^{4}\frac{d^4q_i}{(2\pi)^4}\right)
G^{abcd}_{ \mu \nu \kappa \lambda} 
(q_1, \ldots , q_4) A^{a \mu} (q_1) A^{b \nu} (q_2) A^{c \kappa} (q_3)
 A^{d \lambda} (q_4)(2 \pi)^4 \: \delta^4 (\sum_{i=1}^4 q_i )\,,  
\eqa 
where $V^{abc}_{\nu \kappa \lambda }(q_1,q_2,q_3)$ and
$G^{abcd}_{ \mu \nu \kappa \lambda}(q_1,q_2,q_3,q_4)$ are the usual
three- and four-point vertices in QCD~\cite{Peskin}.


For $n$-gluon scattering the source is given by

\bq
J_{\mu}^{a}=(2 \pi)^4 \, \sum_{i=1}^{n} \epsilon_\mu(p_i,\lambda_i) \delta^{a a_i}
 \delta(q-p_i)
\eq
where $\{ p^\mu_i, \lambda_i,a_i\}$ are the momenta, helicities and colours of the
external gluons. All momenta are taken to be incoming.
The field equations aquire a solution of the form
\bq
A_\mu^a(q)= (2 \pi)^4 \sum_{m=1}^{2^n-2} b_\mu^a(m) \delta(q-P_m)
\label{qcd-solution}
\eq
where $P_m$ are as in \eqn{fie-eq}.
The form of the solution, \eqn{qcd-solution},
suggests that a natural ordering of the momenta can be done in the
following way: for the number $m$, lying between $1$ and $2^n-2$ 
we take its binary
representation, which is an $n$-dimensional vector, $\{m_i\}$,
$i=1,\ldots,n$, with entries 
either $0$ or $1$, and we define 
\[
P^\mu_m = \sum_{i=1}^{n} m_i p_i^\mu\,.
\]
In the usual perturbative sense, one performs an expansion in terms of the
coupling constant
\bq
b_\mu^a(m)=\sum_{k=1}^{n-1} b_\mu^a(m,k) g^{k-1}
\eq
and the field equations take the form:
\bqa
b_\mu^a(m,k) &=& \frac{i}{2 P_m^2} \sum \delta_{m|m_1m_2}
\delta_{k|k_1k_2} V^{aa_1a_2}_{\mu\mu_1\mu_2}(P_m,P_{m_1},P_{m_2})
b^{a_1}_{\mu_1}(m_1,k_1) b^{a_2}_{\mu_2}(m_2,k_2) \nn \\
&-&\frac{1}{6 P_m^2} \sum 
\delta_{m|m_1m_2m_3}
\delta_{k|k_1k_2k_3} G^{aa_1a_2a_3}_{\mu\mu_1\mu_2\mu_3}
b^{a_1}_{\mu_1}(m_1,k_1) b^{a_2}_{\mu_2}(m_2,k_2) b^{a_3}_{\mu_3}(m_3,k_3)
\,,\nn\\
\label{eq-1}
\eqa
where summation over repeated indices is implied and
\[ \delta_{i|i_1i_2}=\delta_{i,i_1+i_2}\,.
\] 
The initial values for the iteration are given by the source term 
and read
\bqa 
b^a_\mu(m,1) =  \left \{ \begin{array}{ll} 
                             \epsilon_\mu(P_m,\lambda_m) \delta^{aa_m} 
                             & m=2^{i-1}, i=1,\ldots,n \\ 
                        0                             & \mbox{otherwise}
                                           \end{array} 
                                        \right.  
\eqa 
\par 

The amplitude is then given by
\bq
{\cal A}=\sum_{a} b^a(1,1)\cdot \hat{b}^a(2^n-2,n-1)
\eq
where the $\hat{}$ means that the propagator factor $1/P_{2^n-2}^2$ has been
removed.

One can further simplify \eqn{eq-1} by defining a Lorentz scalar as
\bq
b^a(m,k,\lambda) \equiv \epsilon^\mu(P_m,\lambda) b_\mu^a(m,k)
\label{scalar}
\eq
where $\lambda$ stands for the helicity of the particle and takes the values
$\lambda=\pm,\mbox{L}$, for transversally and longitudinally polarized gluons.
\eqn{scalar} can be inverted with the solution
\bq
b_\mu^a(m,k) = \sum_{\lambda} \epsilon_\mu(P_m,\lambda) b^a(m,k,\lambda) +
P_{m\mu} \frac{ b^a(m,k)\cdot P_m}{P_m^2}\,.
\eq
Ward identities guarantee that all terms proportional to $P_{m}$ drop out 
in the calculation of the physical amplitude, so that \eqn{eq-1} can be written as

\bqa
b^a(m,k,\lambda) &=& \frac{i}{2 P_m^2} \sum \delta_{m|m_1m_2}
\delta_{k|k_1k_2} \hat{V}^{aa_1a_2}(P_m,\lambda;,P_{m_1},\lambda_1;P_{m_2},\lambda_2)
\nn \\
&&
b^{a_1}(m_1,k_1,\lambda_1) b^{a_2}(m_2,k_2,\lambda_2) \nn \\
&-&\frac{1}{6 P_m^2} \sum 
\delta_{m|m_1m_2m_3}
\delta_{k|k_1k_2k_3} \hat{G}^{aa_1a_2a_3}(\lambda,\lambda_1,\lambda_2,\lambda_3)
\nn \\
&&
b^{a_1}(m_1,k_1,\lambda_1) b^{a_2}(m_2,k_2,\lambda_2) b^{a_3}(m_3,k_3,\lambda_3)
\label{eq-2}
\eqa
where
\bq 
\hat{V}^{abc}(P_i,\lambda_i;P_j,\lambda_j;P_k,\lambda_k) = 
\epsilon^\mu(P_i,\lambda_i) \epsilon^\nu(P_j,\lambda_j) \epsilon^\kappa(P_k,\lambda_k)
V^{abc}_{\mu \nu \kappa } (P_i,P_j,P_k) 
\label{hat3v}
\eq
and
\bq  
\hat{G}^{abcd}(\lambda_1,\lambda_2,\lambda_3,\lambda_4)=
\epsilon^\mu(P_1,\lambda_1) \epsilon^\nu(P_2,\lambda_2) \epsilon^\kappa(P_3,\lambda_3)
\epsilon^\rho(P_4,\lambda_4)
G^{abcd}_{\mu \nu \kappa \rho} 
\eq 

Experience with scalar theory has shown that the computational cost grows like
$~3^n$ for a $\phi^3$ theory and $~4^n$ for a $\phi^4$ theory~\cite{Bru}.
 This suggests that
the computational cost will be lowered if we succeed in writing
 all equations in
terms of three-vertex interactions. This can be done by introducing 
an auxiliary
field $X$ and replacing the four-gluon part of the Lagrangian with 

\bq 
{\cal L}_X= - g f^{abc} A_{ \mu}^{a} A_{ \nu}^{b} X^{ \mu \nu \: c}
- X_{ \mu \nu }^{a}  X^{ \mu \nu \: a }\,. 
\eq 
Then \eqn{eq-2} is written as  
\bqa
b^a(m,k,\lambda) &=& \frac{i}{2 P_m^2} \sum \delta_{m|m_1m_2}
\delta_{k|k_1k_2} \hat{V}^{aa_1a_2}(P_m,\lambda;,P_{m_1},\lambda_1;P_{m_2},\lambda_2)
\nn \\
&&
b^{a_1}(m_1,k_1,\lambda_1) b^{a_2}(m_2,k_2,\lambda_2) \nn \\
&-&\frac{1}{P_m^2} \sum 
\delta_{m|m_1m_2}
\delta_{k|k_1k_2} H^{\mu\nu}(P_m,\lambda;P_{m_1},\lambda_1) f^{aa_1a_2}
\nn \\
&&
b^{a_1}(m_1,k_1,\lambda_1) X^{a_2}_{\mu\nu}(m_2,k_2)
\nn \\
X^a_{\mu\nu}(m,k)&=&-\frac{1}{4}\sum
\delta_{m|m_1m_2} \delta_{k|k_1k_2} 
H_{\mu\nu}(P_{m_1},\lambda_1;P_{m_2},\lambda_2) f^{aa_1a_2}
\nn \\
&&
b^{a_1}(m_1,k_1,\lambda_1) b^{a_2}(m_2,k_2,\lambda_2) 
\label{eq-3}
\eqa
where
\bq
H_{\mu\nu}(p_1,\lambda_1;p_2,\lambda_2)=\left(
\epsilon_\mu(p_1,\lambda_1)\epsilon_\nu(p_2,\lambda_2) -
\epsilon_\nu(p_1,\lambda_1)\epsilon_\mu(p_2,\lambda_2) \right)\,.
\eq


\subsection{ Colour and helicity treatment }  

\eqn{eq-3} can be used to compute the $n$-gluon amplitude for an arbitrary
momentum, colour and helicity configuration of the external particles.
In order to have an estimate of the production probability, one has to
sum over all colour and helicity configurations.
Summation over colours is a delicate subject. If one performs the summation 
in a straightforward way then one has to consider something like
$8^n$ configurations for the $n$-gluon scattering. 
In this section we show how this summation can be replaced by 
integration, which is
then suitable for Monte Carlo computation.

As a first step a simplification of the colour structure of  \eqn{eq-3}
is possible by defining the following object

\bq
b_{AB}\equiv \sum_{a=1}^{8} t^a_{AB} b^a, \;\;\; A,B=1,2,3
\eq
where all other indices have been temporarily suppressed. The new objects
are of course traceless three by three matrices in colour space. The 
interesting property of this
colour representation is that it leads to a `diagonalization'
of the colour structure of the three-gluon vertex. More specifically, the
colour part of the three-gluon vertex is now given by
\bq
f^{abc}t^a_{AB}t^b_{CD}t^c_{EF}=
- \frac{i}{4} (\delta_{A D} \: \delta_{C F} \: \delta_{E B} 
- \delta_{A F} \: \delta_{C B} \: \delta_{E D}) 
\label{diag-col}
\eq

This colour structure is related to the colour flow occuring in the real physical
process, where gluons can be represented by quark-anti-quark states in colour
space and their self-interaction, as given by \eqn{diag-col}, reflects 
the fact that
colour remains unchanged on an uninterrupted colour line.

Accordingly  \eqn{eq-3} is now transformed to 
\bqa
b_{AB}(m,k,\lambda) &=& \frac{1}{2 P_m^2} \sum \delta_{m|m_1m_2}
\delta_{k|k_1k_2} \hat{V}(P_m,\lambda;,P_{m_1},\lambda_1;P_{m_2},\lambda_2)
\nn \\
&&
\left( b_{AC}(m_1,k_1,\lambda_1) b_{CB}(m_2,k_2,\lambda_2)
      -b_{AC}(m_2,k_2,\lambda_2) b_{CB}(m_1,k_1,\lambda_1) \right)
\nn \\
&+&\frac{ i}{P_m^2} \sum 
\delta_{m|m_1m_2}
\delta_{k|k_1k_2} H_{\mu\nu}(P_m,\lambda;P_{m_1},\lambda_1) 
\nn \\
&&
\left( b_{AC}(m_1,k_1,\lambda_1) X_{CB}^{\mu\nu}(m_2,k_2)
      -X_{AC}^{\mu\nu}(m_2,k_2)  b_{CB}(m_1,k_1,\lambda_1) \right)
\nn \\
X_{AB}^{\mu\nu}(m,k)&=&\frac{i}{4}\sum
\delta_{m|m_1m_2} \delta_{k|k_1k_2} 
H^{\mu\nu}(P_{m_1},\lambda_1;P_{m_2},\lambda_2) 
\nn \\
&&
\left( b_{AC}(m_1,k_1,\lambda_1) b_{CB}(m_2,k_2,\lambda_2)
      -b_{AC}(m_2,k_2,\lambda_2) b_{CB}(m_1,k_1,\lambda_1) \right)
\nn\\
\label{eq-4}
\eqa
where 
\[
\hat{V}(P_m,\lambda;,P_{m_1},\lambda_1;P_{m_2},\lambda_2)
\]
stands for the momentum part of the three-gluon vertex, defined in
\eqn{hat3v}.

The next step is to consider a colour representation so that summation 
over colours can be replaced by integration. We seek therefore a colour vector
$\eta^a(z)$ satisfying the following property:
\bq
\int [dz]\, \eta^a(z)\eta^b(z) = \delta^{ab}
\label{eta-norm}
\eq
with some properly defined measure $[dz]$.
This can be realized by standard group-theoretical constructions. In the case
of $SU(3)$, we have to consider the 5-dimensional sphere~\cite{Beg}
parametrized by three complex numbers $(z_1,z_2,z_3)$ subject to the constraint : 
\bq 
\mid z_1 \mid^{2}+\mid z_2 \mid^2+\mid z_3 \mid^2 =1 
\eq 
A useful parametrization of these three complex coordinates is as follows :
\bqa 
&z_1=e^{i \phi_1} \cos \theta & \\ 
&z_2=e^{i \phi_2} \sin \theta \cos \xi &  \\ 
&z_3=e^{i \phi_3} \sin \theta \sin \xi &  \\ \nn 
&0 \leq \phi_i \leq 2 \pi  \; \; ; \; \; 0 \leq \theta \leq \frac{ \pi }{2}  
 \; \; ; \; \;    0 \leq \xi \leq  
\frac{ \pi }{2} &
\eqa 
and the group invariant measure is simply given by
\bq
\int \left( \prod_{i=1}^{3}dz_i dz_i^*\right)  \delta( \sum_{i=1}^{3} z_i z_i^* - 1)
\eq
or, in terms of the polar variables,
\bq 
\left( \prod_{i=1}^{3} \int_0^{2 \pi} d\phi_i \right)
\int_0^{ \frac{ \pi }{2} } d \theta 
\int_0^{ \frac{ \pi }{2} } d \xi 
\cos \theta \sin^3 \theta \cos \xi \sin \xi 
\eq 

It is now a matter of algebra to show that the vectors
\bq 
\eta^{a} (z)= \sqrt{24} \sum_{i,j=1}^{3} z_i^{*} ( t^{a} )_{ij} z_j   
\; \; \; \; \; a=1,\ldots,8 
\eq 
satisfy \eqn{eta-norm}.

As far as our recursive equations are concerned their structure remains 
unaffected
and the only things to be changed are the initial conditions :

\bq
b_{AB}(m,1,\lambda)=\sum_{a=1}^{8}b^a(m,1,\lambda) \eta^a(z)
=\delta_{\lambda\lambda_m} \sqrt{6} \left(
z_{mA} z^*_{mB}-\frac{1}{3} \delta_{AB} \right)
\eq
where as usual $m=2^{i-1},i=1,\ldots,n$, $\lambda_m$ is the helicity
and $\vec{z}_m$ are the new continuous colour 
coordinates of the $m$-th gluon.

In the same way, summation over helicity configurations of the external
gluons can be replaced by an integration over a phase variable. This
is achieved by introducing the polarization vector:

\bq
\epsilon_\phi^\mu(p)=\sqrt{2}\left(
e^{i\phi} \epsilon^\mu(p,+)+e^{-i\phi} \epsilon^\mu(p,-)\right)\,.
\eq

Then by integrating over $\phi$ we obtain the sum over helicities,
\bq
\frac{1}{\pi}\int_{0}^{\pi}d\phi\,\epsilon_\phi^\mu(p) 
(\epsilon_\phi^\nu(p))^*=
\sum_{\lambda=\pm} \epsilon^\mu(p,\lambda) (\epsilon^\nu(p,\lambda))^*\,.
\eq

\section{ Results and Discussion }

Using the recursive equations described so far, we study
the processes $gg\to (n-2)g$, for up to $n=9$. 
The Monte Carlo integration techniques used to represent the colour and 
helicity summation are expected to have an influence 
on the variance of the squared
matrix element over the extended phase space, which now includes besides
the momenta, the continuous colour variables, $z_i,i=1,2,3$, and the helicity 
phase
$\phi$. In order to study the variance we have used the
following computational schemes:
\ben
\item By method {\tt I}
we mean that 
colour and helicity are treated as described in the previous section.
\item
In method {\tt II} the helicity summation is replaced by a discrete
Monte Carlo: in each `event' we randomly choose a helicity configuration
and then multiply by the appropriate combinatorial factor counting the total
number of non-vanishing helicity configurations.
\item
Finally in method {\tt III} a full summation over helicity configurations
has been performed. 
\een
All three methods have been used in the calculation of
 3 and 4 gluon production.
Moreover an `analytic' approximation, the so-called {\tt SPHEL} 
approximation~\cite{sphel,Par-Tay}, has been used in the computation.

In order to avoid collinear and soft singularities the following cuts have
been used:
\bq
p_{Ti} > 60 \mbox{GeV},\;\;\; |\eta_i| < 2,\;\;\; \theta_{ij} > 40^o 
\label{cuts}
\eq

For each given multiplicity, i.e. for a given 
value of $n$, we use a different
centre-of-mass energy, $\sqrt{\hat{s}}$, since higher final-state gluon
multiplicities exhibit higher energy thresholds, due to the
imposed cuts. The computation within
a given gluon multiplicity utilizes the same set of momenta configurations,
so a positive correlation is expected. On the other hand, the variance
can be used to estimate directly the computational speed. As far as the CPU 
time is concerned, methods {\tt I} and {\tt II} use the same amount of
resources, whereas in method {\tt III} we need 10, 
for $n=5$, and 15, for $n=6$,
times more CPU time.
Finally the running strong coupling constant 
with $N_f=5$,  
$\Lambda_{QCD}=200$ MeV and a scale $Q=p_{T,max}$ has been used.

In \tab{t1} the total partonic cross section and the corresponding variance
estimate are presented. We see that
method {\tt I} is generally more efficient than method {\tt II}
since the variance is always better. Moreover, it is also better than method
{\tt III} if we take into account the mulitplicative 
factor 10 ($n=5$) and 15 ($n=6$) needed for the computation. 
Colour and helicity integration do have a 
contribution to the variance as this can be inferred from the comparison
with the variance computed in the case of {\tt SPHEL}, where only phase-space
integration is involved. Of course variance reduction schemes are possible
but their study goes beyond the scope of the present publication.

\begin{table}[th]
\begin{center}
\begin{tabular}{ccccc}
\parbox[c]{2cm}{process\\ $\sqrt{\hat{s}}$}
& {\tt I} 
& {\tt II} 
& {\tt III} 
& {\tt SPHEL} 
\\[12pt]
\parbox[c]{2cm}{$2\to3$\\
                400 GeV}       
& 
\parbox[c]{3cm}{$2.43\pm 0.18$\\
                  $0.034 \pm 0.01$}
& 
\parbox[c]{3cm}{$2.15\pm 0.19$\\ 
                  $0.036 \pm 0.01$}
& 
\parbox[c]{3cm}{$2.35\pm 0.1$\\ 
                  $0.012 \pm 0.002$}
& 
\parbox[c]{3cm}{$2.43\pm 0.06$\\ 
                  $0.004 \pm 0.0003$}
\\[12pt]  
\parbox[c]{2cm}{$2\to4$\\       
                600 GeV}       
& 
\parbox[c]{3cm}{$0.30\pm 0.02$\\
                  $ (4 \pm 1) 10^{-4}$}
& 
\parbox[c]{3cm}{$0.24\pm 0.03$\\ 
                  $ (8 \pm 3) 10^{-4} $}
& 
\parbox[c]{3cm}{$0.31\pm 0.02$\\ 
                  $ (2 \pm 0.5) 10^{-4}$}
& 
\parbox[c]{3cm}{$0.40\pm 0.01$\\ 
                  $ (2 \pm 0.3)10^{-4}$}
\\[12pt]  
\parbox[c]{2cm}{$2\to5$\\       
                900 GeV}       
& 
\parbox[c]{3cm}{$(4.7 \pm 0.3)\, 10^{-2}$\\
                $(  9 \pm 2  )\, 10^{-6}$}
& 
& 
& 
\parbox[c]{3cm}{$(7.6\pm 0.2)\, 10^{-2}$\\
                $(  6\pm 2)\, 10^{-6}  $}
\\[12pt]  
\parbox[c]{2cm}{$2\to6$\\       
                1200 GeV}       
& 
\parbox[c]{3cm}{$(1.0 \pm 0.1)\, 10^{-2}$\\
                $(1.0 \pm 0.4)\, 10^{-6}$}
& 
& 
& 
\parbox[c]{3cm}{$(1.9 \pm 0.1)\, 10^{-2}$\\
                $(1.0 \pm 0.3)\, 10^{-6}$}
\\[12pt]  
\parbox[c]{2cm}{$2\to7$\\       
                1500 GeV}       
& 
\parbox[c]{3cm}{$(1.5 \pm 0.1)\, 10^{-3}$\\
                $(2.1 \pm 0.9)\, 10^{-8}$}
& 
& 
& 
\parbox[c]{3cm}{$(2.3 \pm 0.1)\, 10^{-3}$\\
                $(1.3 \pm 0.3)\, 10^{-8}$}
\\[12pt]  
\end{tabular}
\caption[.]{Total partonic cross section (in nb) and the corresponding
variance with their Monte Carlo errors.}
\label{t1}
\end{center}
\end{table}
\begin{figure}[t]
\begin{center}
\mbox{
\epsfig{file=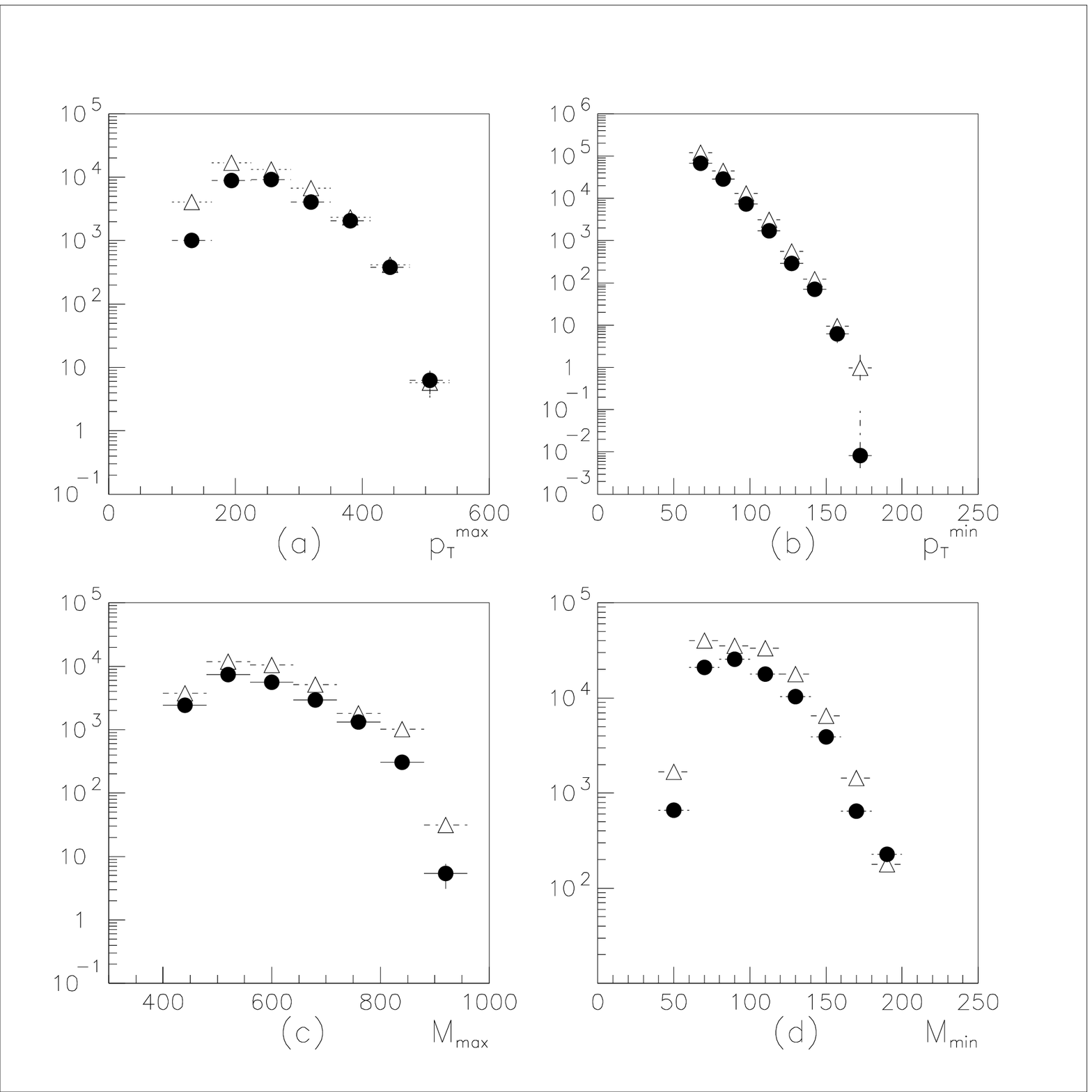,height=8cm,width=8cm}}
\caption[.]{
Differential distributions for the process $gg\to 6g$ in pb/GeV; 
(a) maximum transverse momentum, (b) minimum transverse momentum,
(c) maximum two-jet invariant mass and (d) minimum two-jet invariant mass.
Black circles correspond to the exact result whereas triangles 
represent the predictions of {\tt SPHEL}.}
\label{fig1}
\end{center}
\end{figure}
\begin{figure}[t]
\begin{center}
\mbox{
\epsfig{file=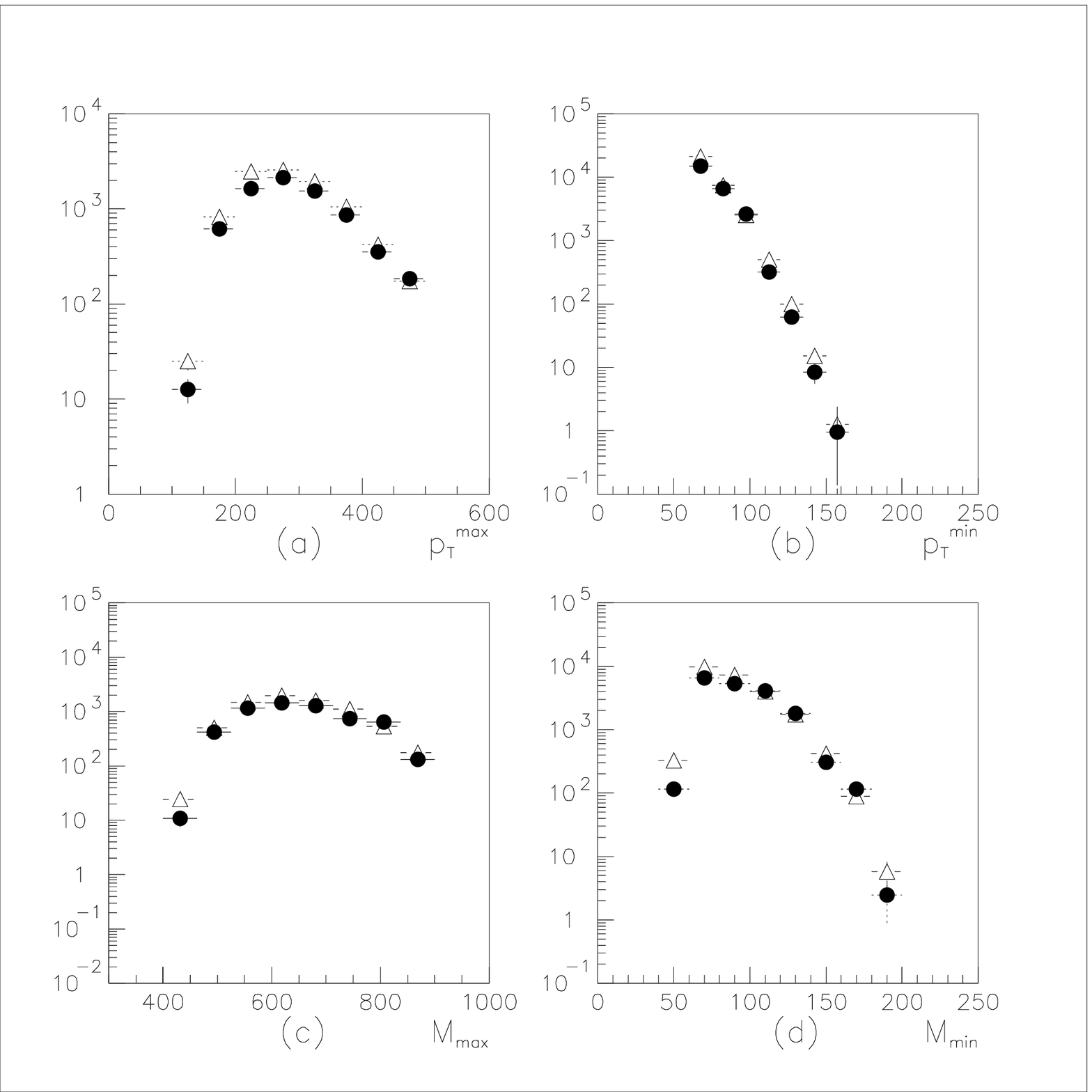,height=8cm,width=8cm}}
\caption[.]{Same as \fig{fig1} for $gg\to 7g$.}
\label{fig2}
\end{center}
\end{figure}
 As a further step we have also considered differential distributions, such as
the maximum and minimum gluon transverse momentum, $p_{T,max}$, $p_{T,min}$,
and the maximum and minimum invariant two-(gluons)jet mass.
In \fig{fig1} the four differential distributions are presented for the
process $gg\to 6g$, whereas in \fig{fig2} the same distributions
are presented for the reaction $gg\to 7g$. In all cases a qualitative agreement
with {\tt SPHEL} has been found. Note, however, that our computational
scheme is an exact tree-order calculation of the amplitude, where no
leading colour approximation nor special helicities selection 
have been used. 

In summary a computational method based on recursive equations has been
presented which enables one to compute the amplitudes for the processes
$gg\to (n-2)g$ for at least up to $n=9$. The method can be easily extended
to include all partonic subprocesses~\cite{drag_thesis}. 
It can provide a reliable tree-order
event-generator for multi-jet production at LHC energies.

\end{document}